\documentclass[12pt]{iopart}

\usepackage{iopams}
\usepackage{amssymb}
\usepackage{bm}
\usepackage{amsfonts}
\usepackage{graphicx,epsfig}
\usepackage{graphicx,color}

\begin{document}

\title{Current biased gradiometric flux qubit in a circuit-QED architecture}

\author{Mun Dae Kim}

\address{College of Liberal Arts, Hongik University, Sejong 30016, Korea}
\ead{mundkim@gmail.com}
\vspace{10pt}


%
%
%
%

\begin{abstract}
We propose a scheme for controlling  the gradiometric flux qubit
(GFQ) by applying an ac bias current in a circuit-QED
architecture. The GFQ is insensitive to the magnetic flux
fluctuations, which at the same time makes it challenging to
manipulate the qubit states by an external magnetic field. In this
study, we demonstrate that  an ac bias current applied to the
$\alpha$-junction of the GFQ can control the qubit states.
Further, the present scheme is robust against  the charge
fluctuation as well as the magnetic flux fluctuations, promising a
long coherence time for quantum gate operations. We introduce a
circuit-QED architecture  to perform the single and two-qubit
operations with a sufficiently strong coupling strength.
\end{abstract}

\noindent{\it Keywords\/}: gradiometric flux qubit, ac current bias, circuit-QED

\section{Introduction}

Superconducting qubits are promising candidates for noisy intermediate-scale quantum(NISQ)
computing due to the scalability and  flexibility of circuit design.
Unfortunately the electrical and magnetic noises in charge- and
phase-based qubits, respectively, are the severe decoherence
sources.  While the fluctuations of magnetic fluxes in preparing
and manipulating the quantum states of the superconducting flux
qubits deteriorate the qubit state coherence, for charge qubits
the qubit state suffers from the charge fluctuations. The transmon
qubit  which is coupled with the transmission line resonator at a
sweet spot of gate voltage is robust against the charge
fluctuations \cite{Koch}. On the other hand, for phase-based
qubit, the gradiometric flux qubit(GFQ) has been intensively
investigated because GFQ is  robust against the
flux fluctuations due to the symmetric design
\cite{Paauw,Fedorov10,Fedorov11,Schwarz,Wang17,Wang19}.

The simplest three-Josephson junction flux qubit has a fixed gap
which depends on the critical current  of the so-called
$\alpha$-junction \cite{Orlando}. The  $\alpha$-junction has a
reduced critical current, $I_{c,\alpha}=\alpha I_c$, which can be
replaced by a dc-SQUID loop in order to {\it in situ} adjust the
critical current of the $\alpha$-junction and thus the qubit
energy gap $\Delta$. This type of tunable-gap flux qubit, however,
is sensitive to the magnetic flux fluctuations responsible for the
short decoherence time of qubit operations. On the other hand, the
GFQ in Fig. \ref{scheme} has a symmetric structure so that the
magnetic fluxes $f_1$ and $f_2$, where $f_i=\Phi_{{\rm
ext},i}/\Phi_0$ with the external magnetic flux  $\Phi_{{\rm
ext},i}$ and the unit flux quantum $\Phi_0=h/2e$, become
compensated in the left and right trapping loops, and thus the
global magnetic noises cannot affect the qubit states  \cite{Schwarz}.
The GFQ is robust against such global noises
that  originate from  the residual thermal photons in the resonator 
which is the dominant dephasing source at the symmetric point \cite{Yan}.

 However, at the same time,  if we apply a magnetic field to manipulate the
qubit states, the qubit also  does not respond to the operating
flux. In order to control the qubit state one should	 apply a magnetic
energy bias which can be induced by generating an asymmetry,
$\epsilon=f_1-f_2$, in magnetic fluxes threading into the left and
right trapping loop of the GFQ \cite{Paauw,Fedorov10,Fedorov11,Schwarz},
resulting in the deviation of the GFQ state  from the symmetric point.
This asymmetry then may cause a decoherence to the GFQ states.
The $1/f$ noise from local magnetic defects in the qubit loop 
may be detrimental to the flux qubit coherence. 
However, the flux qubit is well protected from the $1/f$ noise at its symmetric point \cite{Clarke}. 
In the present study we introduce an ac bias current scheme for operating the GFQ state 
at the symmetric point, diminishing the  decoherence of the GFQ states due to the local defects.

In recently studied schemes  an ac bias current is applied to the
three-Josephson junction flux qubit to generate a strong coupling
between a superconducting qubit and resonator
\cite{Kim10,Steffen,Inomata,Inomata14,Zagoskin,Kim15,Kim17}, a
two-qubit coupling \cite{Chow,Strand},  and a resonator-resonator
coupling \cite{Baust15,Baust16,Wulschner}. In those schemes the
asymmetry of three-Josephson junction qubit is the origin of the
current-qubit coupling. However, this asymmetry may also give rise to
the decoherence due to the magnetic flux fluctuations. Hence in
this study we introduce an ac current bias scheme for the
symmetric GFQ in a circuit-quantum electrodynamics(QED)
architecture.
The quantized voltage and current modes in the transmission line
resonator interact with the qubit coupled to the circuit QED.
While charge-based qubits such as transmon are coupled with the
voltage modes of  the resonator, usually flux qubits are
inductively coupled with the current modes through the mutual
inductance between the qubit loop and resonator
\cite{Lindstrom,Oelsner}. Previously a  GFQ has been coupled to
the resonator of circuit-QED architecture via the mutual
inductance, providing a weak coupling strength \cite{Wang17}.

In this study we provide a scheme for  the GFQ biased by an
ac current mode of circuit-QED resonator. 
By obtaining the exact Lagrangian of the system analytically we show  that the
$\alpha$-junction loop of the  GFQ can be biased by the ac current,
$I_0$, in Fig. \ref{scheme}(a) to perform the quantum gate operations.
However,  the phases,  $\varphi_i (i\!=\!1\!\sim\! 4)$,
of Josephson junctions in the
GFQ altogether can not be coupled to the bias current, $I'_0$,
which means that the GFQ does not respond to the
current noises originating from the external charge fluctuations  with scale larger than size of the GFQ.
The GFQ is insensitive to flux fluctuation, and moreover
the ac current biased GFQ is also  robust against the charge
fluctuations as well, which will guarantee a long coherence time
enough  for the quantum gate operations. Furthermore, the present ac current
bias scheme can induce a sufficiently strong GFQ-resonator
coupling to perform the single qubit and two-qubit gate operation
in a circuit-QED architecture.

\section{Ac current biased gradiometric flux qubit}

\subsection{Effective potential for ac bias current coupling}

\begin{figure}[t]
\vspace{-1cm}
\centering
\includegraphics[width=1.0\linewidth]{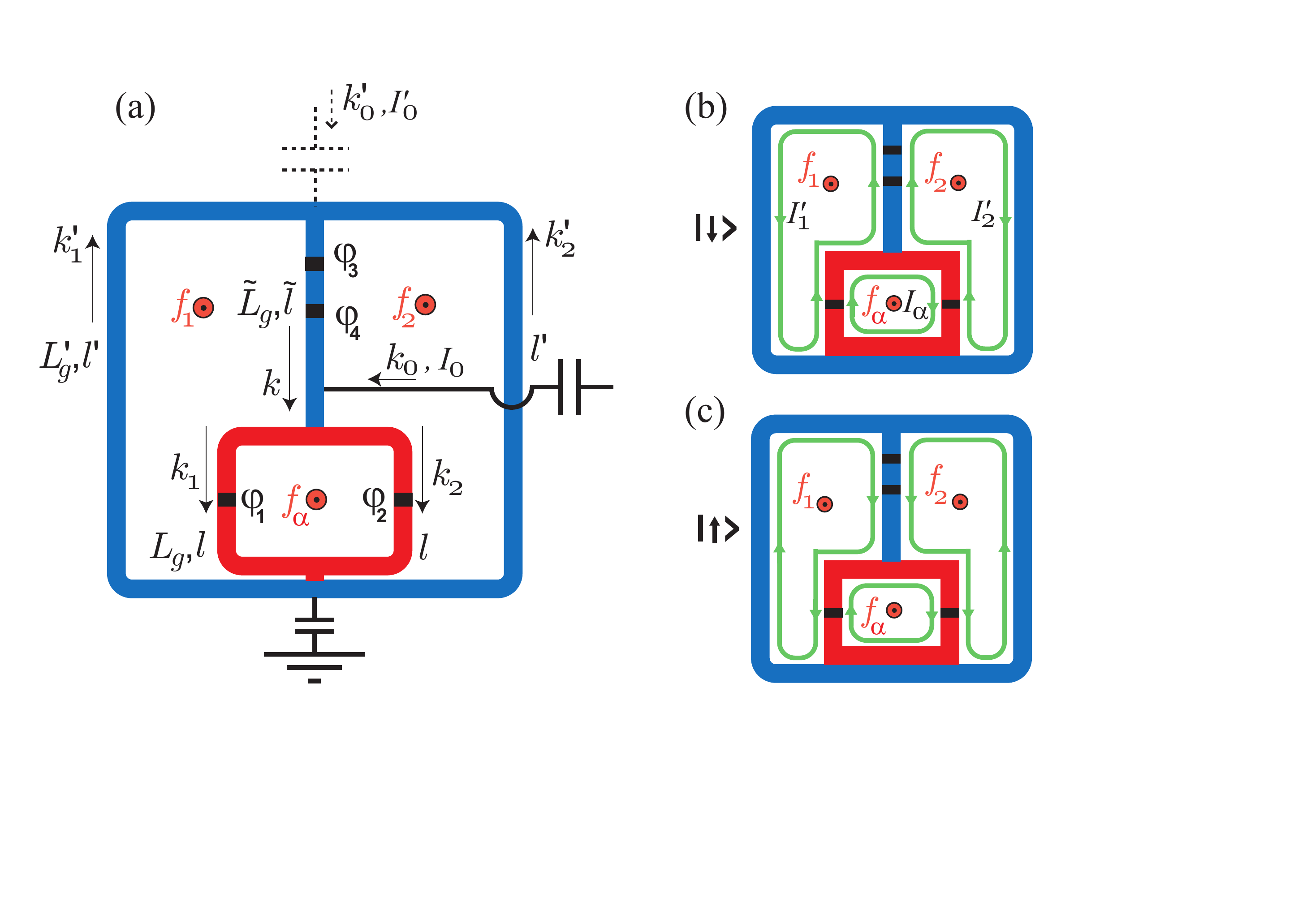}
\vspace{-3cm} \caption{(a)A scheme of gradiometric flux qubit with
an ac bias current $I_0$ which can control the qubit state while
the ac bias current $I'_0$ through the dotted line in the figure
cannot affect the qubit state. The central dc-SQUID takes the role
of the $\alpha$-junction loop which provides  tunable qubit energy
gap $\Delta(f_\alpha)$. (b) and (c) show the currents of the qubit
states $|\downarrow\rangle$ and $|\uparrow\rangle$, respectively,
where  the directions of the trapping loop currents are opposite
with each other.
}
\label{scheme}
\end{figure}

In Fig. \ref{scheme}(a) we introduce an  ac current bias  scheme
for manipulating the GFQ state consisting of left and right
trapping loops and $\alpha$-junction loop with threading magnetic
fluxes, $\Phi_{\rm ext,i}$ and $\Phi_{\rm ext,\alpha}$,
respectively. Here we consider that  an  ac bias current $I_0$ is
applied to the $\alpha$-junction loop.
In order to obtain the effective potential describing the dynamics
of the system, we consider  periodic boundary conditions
originating from the usual fluxoid quantization condition of
superconducting loop \cite{Tinkham,KimSR}. In the system of Fig.
\ref{scheme}(a) we have three independent loops whose boundary
conditions can be given by
\begin{eqnarray}
\label{bc1}
&-& k_1 l - k'_1 l'-k{\tilde l}-\varphi_1-\varphi_3-\varphi_4=2\pi(n_1+f_1+f_{1,{\rm ind}})\\
\label{bc2}
&&k_2 l+k'_2 l'+k{\tilde l}+\varphi_2+\varphi_3+\varphi_4=2\pi(n_2+f_2+f_{2,{\rm ind}})\\
\label{bc3}
&&k_1l-k_2l+\varphi_1-\varphi_2=2\pi(n+f_\alpha+f_{\alpha,{\rm ind}}),
\end{eqnarray}
where $k_i$, ${\tilde l}$, $l'$ and $l$ are the wave vector of the
Cooper pairs, the lengths of the central branch, the left(right)
branch, and half the length of  the dc-SQUID loop, respectively.
Here, $\varphi_i$'s are the phase differences of the Cooper pair
wave function across the Josephson junction, $f_{1(2),{\rm ind}}$
and $f_{\alpha, {\rm ind}}$ are the induced magnetic fluxes of the
left(right) trapping loop and the dc-SQUID loop, and  $n_i$'s  are
integer. Here, the dc-SQUID loop has the role of the
$\alpha$-junction loop which provides the tunable qubit energy gap
$\Delta(f_\alpha)$.

The induced fluxes, $f_{\rm ind,i}=\Phi_{\rm ind,i}/\Phi_0$, are
given by $f_{\rm ind,1}=(1/\Phi_0)(-L_gI_1-{\tilde L}_gI-L'_gI'_1-L_MI_2-L_M'I'_2)$,
$f_{\rm ind,2}=(1/\Phi_0)(L_gI_2+{\tilde L}_gI+L'_gI'_2+L_MI_1+L_M'I'_1)$, and
$f_{\rm ind,\alpha}=(1/\Phi_0)(L_gI_1-L_gI_2)$. Here, the Cooper pair
current $I_i$ is represented by $I_i=-(n_cAq_c/m_c)\hbar k_i$ with  the
Cooper pair density $n_c$, the cross section $A$ of the loop,
$q_c=2e$ and $m_c=2m_e$. The induced flux, $\Phi_{\rm ind,1}$, in the left trapping loop, for example,
consists of contributions from the geometric self inductances
$L'_g$, ${\tilde L}_g$ and $L_g$  of the left branch, the central branch and the left
half of  the SQUID loop, respectively, and the geometric mutual inductances $L_M$ and $L_M'$
between the left trapping loop and  the right half of the SQUID loop and the right branch, respectively.
For the induced flux, $f_{\rm ind,\alpha}$, of
$\alpha$-junction loop the contributions of mutual inductances
between the $\alpha$-junction loop and the left and right trapping
loops are cancelled, considering the current directions in Figs. \ref{scheme}(b) and (c).

We also introduce the kinetic inductances $L_K=m_cl/An_cq^2_c$, $L'_K=m_cl'/An_cq^2_c$,
and  ${\tilde L}_K=m_c{\tilde l}/An_cq^2_c$, \cite{Kim04,Meservey,Hazard}
and then the induced fluxes  become
$f_{\rm ind,1}=[(L'_g/L'_K)(l'/2\pi)k'_1+(L_g/L_K)(l/2\pi)k_1+({\tilde L}_g/{\tilde L}_K)({\tilde l}/2\pi)k
+(L_M'/L'_K)(l'/2\pi)k'_2+(L_M/L_K)(l/2\pi)k_2]$,
$f_{\rm ind,2}=-[(L'_g/L'_K)(l'/2\pi)k'_2+(L_g/L_K)(l/2\pi)k_2+({\tilde L}_g/{\tilde L}_K)({\tilde l}/2\pi)k
+(L_M'/L'_K)(l'/2\pi)k'_1+(L_M/L_K)(l/2\pi)k_1]$,
and $f_{\rm ind,\alpha}=-(L_g/L_K)(l/2\pi)(k_1-k_2)$
to represent the boundary conditions as
\begin{eqnarray}
\label{lbc}
\fl &-&\!\!\!\left(\!\!1\!+\!\frac{L_g}{L_K}\!\right)\!k_1 l\!-\!\!\left(\!\!1\!+\!\frac{L'_g}{L'_K}\!\right)\!k'_1 l'\!
-\!\!\left(\!\!1\!+\!\frac{{\tilde L}_g}{{\tilde L}_K}\!\right)\!k {\tilde l}\!-\!\frac{L_M'}{L'_K}k'_2 l'\!-\!\!\frac{L_M}{L_K}k_2 l
\!=\!2\pi\!\left(\!\!n_1\!+\!f_1\!+\!\frac{\varphi_1\!\!+\!\!\varphi_3\!\!+\!\!\varphi_4}{2\pi}\!\!\right)\\
\label{rbc}
\fl &&\!\!\!\left(\!\!1\!+\!\frac{L_g}{L_K}\!\right)\!k_2 l\!
+\!\!\left(\!\!1\!+\!\frac{L'_g}{L'_K}\!\right)\!k'_2 l'\!+\!\!\left(\!\!1\!+\!\frac{{\tilde L}_g}{{\tilde L}_K}\!\right)\!k {\tilde l}
\!+\!\frac{L_M'}{L'_K}k'_1 l'\!+\!\!\frac{L_M}{L_K}k_1 l
\!=\!2\pi\!\left(\!\!n_2\!+\!f_2\!-\!\frac{\varphi_2\!\!+\!\!\varphi_3\!\!+\!\!\varphi_4}{2\pi}\!\!\right)\\
\label{cbc}
\fl &&\!\!\left(1\!+\!\frac{L_g}{L_K}\right)(k_1\!-\!k_2)l=2\pi\left(n\!+\!f_\alpha\!-\!\frac{\varphi_1\!-\!\varphi_2}{2\pi}\right).
\end{eqnarray}
If we assume that the GFQ loops and bias line have the same cross section $A$ and
Cooper pair density $n_c$, the current conservation conditions
at the nodes of the circuit can be represented as
\begin{eqnarray}
\label{kbc}
k=k'_1+k'_2, ~~k+k_0=k_1+k_2.
\end{eqnarray}

From the boundary conditions in Eqs. (\ref{lbc}), (\ref{rbc}) and
(\ref{cbc}) in conjunction with Eq. (\ref{kbc}) we can obtain the
wave vectors, $k_i$, as follows:
\begin{eqnarray}
\label{ki}
\fl k_{1,2}&\!=\!& \frac{2\pi L_K}{l}\!\left[\frac{1}{2L_{\rm eff}}\!\left(\!m\!+\!f_2\!-\!f_1\!-\!\frac{\varphi_1\!+\!\varphi_2\!+\!2\varphi_3\!+\!2\varphi_4}{2\pi}\right)\!\pm\!\frac{1}{2(L_K\!\!+\!\!L_g)}\left(n\!+\!f_\alpha\!-\!\frac{\varphi_1\!-\!\varphi_2}{2\pi}\right)\right]\nonumber\\
\fl &+&\frac{L'_K\!+\!L'_g\!+\!2({\tilde L}_K\!+\!{\tilde L}_g)\!+\!L_M'}{2L_{\rm eff}}k_0\\
\label{k'i}
\fl k'_{1,2}&\!=\!\!&\frac{2\pi L'_K}{l'}\!\!\left[\!\frac{1}{2L_{\rm eff}}\!\left(\!m\!+\!f_2\!-\!f_1\!-\!\frac{\varphi_1\!+\!\varphi_2\!+\!2\varphi_3\!+\!2\varphi_4}{2\pi}\!\right)\!
\mp\frac{m'\!+\!f_1\!+\!f_2\!+\!(1\!-\!\frac{L_M}{L_k+L_g})f_\alpha}{2(L'_K+L'_g)}\right]\nonumber\\
\fl \!\!&-&\!\frac{L_K\!+\!L_g\!+\!L_M}{2L_{\rm eff}}k_0\\
\label{k}
\fl k\!&=&\frac{2\pi {\tilde L}_K}{{\tilde l}L_{\rm eff}}\left(m+f_2-f_1-\frac{\varphi_1+\varphi_2+2\varphi_3+2\varphi_4}{2\pi}\right)
-\frac{L_K+L_g+L_M}{L_{\rm eff}}k_0
\end{eqnarray}
with $L_{\rm eff}\equiv L_K+L_g+L'_K+L'_g+2({\tilde L}_K\!+\!{\tilde L}_g)+L_M+L_M'$,  $m=n_2-n_1$ and $m'=n_1+n_2$.

By  using the voltage-phase relation $V=-(\Phi_0/2\pi){\dot \phi}$
the current relation for Josephson junction in the capacitively-shunted model
is written as $I=-I_c\sin\phi+C{\dot V}=-I_c\sin\phi-C(\Phi_0/2\pi){\ddot
\phi}$, where  $I_c$ is the critical current and $C$  the capacitance of Josephson
junction.  The quantum Kirchhoff relation then is represented as
$-(\Phi^2_0/2\pi L_K)(l/2\pi)k_i=-E_{J}\sin\phi_i-C(\Phi_0/2\pi)^2{\ddot \phi}_i$
with the Josephson coupling energy  $E_J=\Phi_0I_c/2\pi$ and the
current  $I=-(n_cAq_c/m_c)\hbar k$. The equation of motion of the
phase variables of  a Josephson junction,
$C_i(\Phi_0/2\pi)^2{\ddot\phi}_i=-\partial U_{\rm
eff}/\partial\phi_i$, can be  derived from the Lagrange equation
$(d/dt)\partial {\cal L}/\partial {\dot \phi}_i-\partial{\cal
L}/\partial\phi_i=0$ with  the Lagrangian ${\cal
L}=\sum_i(1/2)C_i(\Phi_0/2\pi)^2{\dot \phi}^2_i-U_{\rm
eff}(\{\phi_i\})$ and the effective potential of the system
$U_{\rm eff}(\{\phi_i\})$. By using this equation of motion the
 quantum Kirchhoff relation becomes
\begin{eqnarray}
\label{QK}
\frac{\Phi^2_0}{2\pi L_K}\frac{l}{2\pi}k_i-E_{J}\sin\phi_i=-\frac{\partial U_{\rm eff}}{\partial\phi_i}
\end{eqnarray}
from which we can find the effective potential $U_{\rm eff}(\{\phi_i\})$  satisfying the quantum Kirchhoff relation
as
\begin{eqnarray}
\label{Ueff}
\fl U_{\rm eff}(\{\phi_i\})&\!=\!&\frac{\Phi^2_0}{4L_{\rm eff}}
\!\left(\!m\!+\!f_2\!-\!f_1\!-\!\frac{\varphi_1\!+\!\varphi_2\!+\!2\varphi_3\!+\!2\varphi_4}{2\pi}\right)^2\!\!
\!\!+\!\!\frac{\Phi^2_0}{4(L_K\!+\!L_g)}\!\left(\!n\!+\!f_\alpha\!-\!\frac{\varphi_1\!-\!\varphi_2}{2\pi}\right)^2\\
\fl &\!-\!&\!\!\sum^4_{i=1}\!E_{Ji}\cos\varphi_i\!+\!\!\frac{\Phi_0I_0}{4\pi L_{\rm eff}}
[(L'_K\!\!+\!\!L'_g\!\!+\!\!2{\tilde L}_K\!\!+\!\!2{\tilde L}_g\!\!+\!\!L_M')(\varphi_1\!\!+\!\!\varphi_2)\!
\!-\!\!2(L_K\!\!+\!\!L_g\!\!+\!\!L_M)(\varphi_3\!\!+\!\!\varphi_4)].\nonumber
\end{eqnarray}
Here the first and second  terms correspond to the inductive energy of the trapping loops
and the $\alpha$-junction loop, the third term the Josephson junction energies,
and the last term the coupling energy between the bias current and the phases.

\begin{figure}[t]
\vspace{0cm}
\hspace{2cm}
\includegraphics[width=1.1\linewidth]{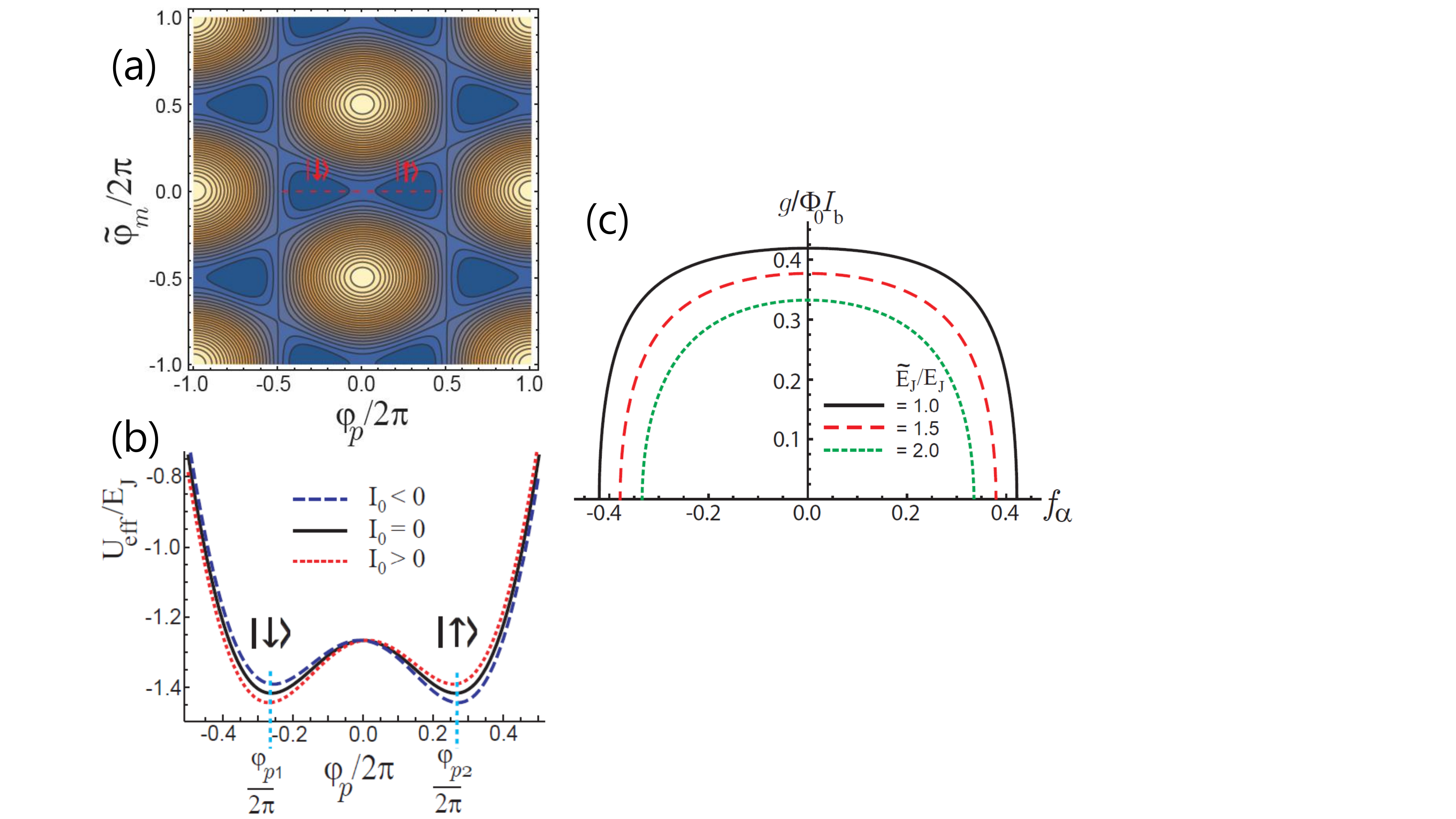}
\vspace{-0.5cm}
\caption{(a) Numerical contour plot for the effective potential in Eq. (\ref{Utr}) for $I_0=0$.
The states at two local minima are denoted as $|\downarrow\rangle$ and $|\uparrow\rangle$.
Here we set $f_\alpha = 0.2$, ${\tilde E}_J /E_J= 2.0$, and $f_1= f_2$.
(b) Double well potential along the dotted line in (a).  For finite bias current $I_0$
the potential becomes tilted in opposite directions depending on the sign of $I_0$.
(c) Coupling strength $g$ between the GFQ and the bias current with amplitude $I_b$. }
\label{contourwell}
\end{figure}

Here, we introduce a coordinate transformation,
$\varphi_{p,m}=(\varphi_1\pm\varphi_2)/2$ and
$\tilde{\varphi}_{p,m}=(\varphi_3\pm\varphi_4)/2$, to represent the
effective potential as
\begin{eqnarray}
\label{Utr}
\fl U_{\rm eff}(\varphi_{p,m},\tilde{\varphi}_{p,m})&=&\frac{\Phi^2_0}{4L_{\rm eff}}\left(m+f_2-f_1
-\frac{2\varphi_p+4{\tilde\varphi}_p}{2\pi}\right)^2
+\frac{\Phi^2_0}{4(L_K+L_g)}\left(n+f_\alpha-\frac{\varphi_m}{\pi}\right)^2 \nonumber\\
\fl &-&2E_J\cos\varphi_p\cos\varphi_m-2{\tilde E}_J\cos{\tilde \varphi}_p\cos{\tilde \varphi}_m \\
\fl &+&\frac{\Phi_0I_0}{2\pi L_{\rm eff}} [(L'_K+L'_g+2{\tilde L}_K+2{\tilde L}_g+L_M')\varphi_p-2(L_K+L_g+L_M){\tilde\varphi}_p],
\nonumber
\end{eqnarray}
where we set $E_{J1}=E_{J2}=E_J$ and $E_{J3}=E_{J4}={\tilde E}_J$.
In Fig. \ref{contourwell}(a) we show the effective
potential $U_{\rm eff}(\varphi_{p,m},\tilde{\varphi}_{p,m})$ as a
function of $\varphi_p$ and ${\tilde \varphi}_m$ for $I_0=0$. The value of
$\varphi_m$ and ${\tilde \varphi}_p$ are determined by numerically
minimizing the effective potential $U_{\rm
eff}(\varphi_{p,m},\tilde{\varphi}_{p,m})$. Actually we have found
that for different  even-odd parity of $m$ and $n$   the local
minima  connected by the dotted line in Fig. \ref{contourwell}(a)
can be formed with the double well structure as shown in Fig.
\ref{contourwell}(b). Here we set  $m=0$, $n=1$,  $f_\alpha =
0.2$, ${\tilde E}_J /E_J= 2.0$, and further $f_1=f_2$ so the external fluxes, $f_1$ and $f_2$, do not
affect the qubit states.

At the  two local minima denoted as $|\uparrow\rangle$ and
$|\downarrow\rangle$ in Fig. \ref{contourwell}(a) we found that
the inductive energy, $U_{\rm ind}$, of the first  two terms in
the effective potential of Eq. (\ref{Utr}) is $U_{\rm ind}/E_J
\sim 0.006$, while the Josephson junction energy, $U_{\rm JJ}$, of
third and forth terms $U_{\rm JJ}/E_J\sim -2.886$. Hence, for
simplicity, we can neglect the induced energies in the effective
potentials of Eqs. (\ref{Ueff}) and (\ref{Utr}). Actually in
experimental situations  the geometric inductance is dominant over the kinetic inductance, $L_g \gg L_K$,
and the inductive energy $\Phi^2_0/L_g \sim O(10^3 E_J)$ \cite{Wal}  is much larger than the Josephson energy scale $E_J$,
which means that  the following boundary conditions should be
satisfied,
\begin{eqnarray}
\label{alphabc}
&& n+f_\alpha-\frac{\varphi_1-\varphi_2}{2\pi}\approx 0,\\
\label{loopbc}
&& m+f_2-f_1-\frac{\varphi_1+\varphi_2+2\varphi_3+2\varphi_4}{2\pi}\approx 0,
\end{eqnarray}
in order to neglect the first two terms of   $U_{\rm eff}(\{\phi_i\})$ in Eq.
(\ref{Ueff})  in spite of the large values of
$\Phi^2_0/4L_{\rm eff}E_J$  and $\Phi^2_0/4(L_K+L_g)E_J$. Here, we
set $\Phi^2_0/4L_{\rm eff}E_J=1000$ and
$\Phi^2_0/4(L_K+L_g)E_J=3000$.
The boundary condition of  Eq. (\ref{alphabc}) can also be obtained
from  Eq. (\ref{bc3}) because in the parameter regime for flux
qubits the phase evolution, $kl$, of Cooper pair wave function and
the induced magnetic flux,  $f_{\alpha,{\rm ind}}$, are negligible
compared to  the phase difference $\varphi$ across the Josephson
junction. In the same way  the boundary condition of Eq.
(\ref{loopbc}) can also be obtained by subtracting two boundary
conditions of Eqs. (\ref{bc1}) and (\ref{bc2}).

We can represent the boundary conditions in Eqs. (\ref{alphabc}) and (\ref{loopbc}) in the transformed coordinate as
$  n+f_\alpha-\varphi_m/\pi\approx 0$ and $m+f_2-f_1-(2\varphi_p+4{\tilde\varphi}_p)/2\pi\approx 0$, and thus
the last term of Eq. (\ref{Utr}) can be written as
\begin{eqnarray}
\fl &&\frac{\Phi_0I_0}{2\pi L_{\rm eff}} [(L'_K+L'_g+2{\tilde L}_K+2{\tilde L}_g+L_M')\varphi_p+(L_K+L_g+L_M)(\varphi_p-\pi(m+f_2-f_1))] \nonumber\\
\fl &=&\frac{\Phi_0I_0}{2\pi}\left[\varphi_p-\frac{(L_K+L_g+L_M)\pi}{L_{\rm eff}}(m+f_2-f_1)\right].
\end{eqnarray}
The effective potential $U_{\rm eff}(\varphi_{p,m},\tilde{\varphi}_{p,m})$ in Eq. (\ref{Utr})
then can be approximated to a function of $\varphi_p$ and ${\tilde \varphi}_m$, $V(\varphi_p,{\tilde \varphi}_m)$,
apart from the constant term:
\begin{eqnarray}
\label{V}
\fl V(\varphi_p,{\tilde \varphi}_m)\!=\! -2E_J\cos\pi(n\!+\!f_\alpha)\cos\varphi_p
\!-\!2{\tilde E}_J\cos\frac{\pi(m\!+\!f_2\!-\!f_1)\!-\!\varphi_p}{2}\cos{\tilde \varphi}_m
\!+\!\frac{\Phi_0 I_0}{2\pi}\varphi_p,
\end{eqnarray}
which shows that the GFQ  can be coupled to the  ac bias current
through only the phase, $\varphi_\alpha$, of Josephson junctions
in the $\alpha$-junction loop with the strength  independent of the values of  individual inductances.
Alternatively, with the boundary condition of Eq. (\ref{loopbc}) the coupling between the bias current and  phase
can also be represented in terms of  ${\tilde\varphi}_p$.
%
%

On the contrary, the phases,  $\varphi_i
(i\!\!=\!\!1\!\!\!\sim\!\!\! 4)$, of Josephson junctions in the
GFQ altogether cannot be coupled to the bias current. Until now
we have considered the bias current $I_0$ in Fig. \ref{scheme}(a).
However, we can also apply the bias current $I'_0=-(n_cAq_c/m_c)\hbar k'_0$,
denoted as dotted line in Fig. \ref{scheme}(a),
at the node that the wave vectors $k, k'_1$ and $k'_2$ meet.
Instead of the boundary conditions in Eq. (\ref{kbc}) in this case
we have the conditions $k'_1+k'_2+k'_0=k$ and $k=k_1+k_2$. Through a similar derivation to the previous one for
Fig. \ref{scheme}(a) with above conditions in conjunction with  Eqs. (\ref{bc1})-(\ref{bc3})
we can obtain the effective potential
\begin{eqnarray}
\label{Ueff1}
\fl U_{\rm eff}(\{\phi_i\})&=&\!\frac{\Phi^2_0}{4L_{\rm eff}}
\left(\!m\!+\!f_2\!-\!f_1\!-\!\frac{\varphi_1\!+\!\varphi_2\!+\!2\varphi_3\!+\!2\varphi_4}{2\pi}\right)^2
\!+\!\frac{\Phi^2_0}{4(L_K+L_g)}\left(\!n\!+\!f_\alpha\!-\!\frac{\varphi_1\!-\!\varphi_2}{2\pi}\right)^2 \nonumber\\
\fl &-&\sum^4_{i=1}E_{Ji}\cos\varphi_i
+\frac{\Phi_0I'_0(L'_K+L'_g+L_M')}{4\pi L_{\rm eff}} (\varphi_1+\varphi_2+2\varphi_3+2\varphi_4),
\end{eqnarray}
where the last term describing the coupling to the bias current
$I'_0$ has a  form different from  that in Eq. (\ref{Ueff}).

Similarly to the  previous case we also have the boundary
condition,
$(\varphi_1+\varphi_2+2\varphi_3+2\varphi_4)/2\pi\approx
m+f_2-f_1$. The last  term of Eq. (\ref{Ueff1}) then becomes a
constant term,  $[\Phi_0I'_0(L'_K+L'_g+L_M')/2L_{\rm eff}] (
m+f_2-f_1)$, which means that with this scheme  the bias-current
cannot be coupled with the phases of the Josephson junctions of
the GFQ and  thus we cannot control the qubit state in contrast to
the previous scheme in Fig. \ref{scheme}(a). Further, the charge
fluctuations with scale larger than size of the GFQ   can be
considered to induce   noisy bias currents through the dotted line
in Fig. \ref{scheme}(a) which, however, cannot affect the qubit
state. Hence  the present ac bias current scheme for the GFQ is robust against the  charge
fluctuations as well as the magnetic flux fluctuations, which may provide a long coherence time for the NISQ
computing.

In Fig. \ref{scheme}(a) the bias current line needs an airbridge structure which is 
not so simple  to construct. Many experimental efforts have been devoted to mitigate the additional loss 
to be small \cite{Chen,Dunsworth}, and moreover the undesired effect of airbridge capacitance 
has also been studied to be kept at a minimum level \cite{Abuwasib}. However, through the small capacitance 
a residual current, ${\tilde I}=-(n_cAq_c/m_c)\hbar {\tilde k}$ with the Cooper pair wave vector ${\tilde k}$, 
may flow into the branch line. The current conservation conditions then  become  $k'_1+k'_2+{\tilde k}=k$ 
and $k=k_1+k_2$, which are the same as those conditions above Eq. (\ref{Ueff1}) 
for the current $I'_0=-(n_cAq_c/m_c)\hbar k'_0$,
and thus gives rise to the same constant coupling term as the last term of Eq. (\ref{Ueff1}). 
Hence, the GFQ state will be also robust against the fluctuations due to the residual current ${\tilde I}$ 
through the airbridge.

%

\subsection{Coupling strength between GFQ and ac bias current }

The effective potential in Eq. (\ref{V}) for $I_0=0$  with  $m=0$, $n=1$ and $f_1=f_2$ becomes
$V(\varphi_p,{\tilde \varphi}_m)= 2E_J\cos(\pi f_\alpha)\cos\varphi_p
-2{\tilde E}_J\cos(\varphi_p/2)\cos{\tilde \varphi}_m$. Here we can obtain analytically
the position of local minima,  $(\varphi_{p1},{\tilde \varphi}_{m1})$ and  $(\varphi_{p2},{\tilde \varphi}_{m2})$,
corresponding to the states  $|\downarrow\rangle$ and $|\uparrow\rangle$ with energy levels
$E_\downarrow$ and $E_\uparrow$, respectively,
which are symmetric with respect to $\varphi_p=0$ as shown in Fig. \ref{contourwell}(a).
From  $\partial V(\varphi_p,{\tilde \varphi}_m)/\partial {\tilde\varphi}_m=
2{\tilde E}_J \cos(\varphi_p/2)\sin{\tilde \varphi}_m=0$
and $\partial V(\varphi_p,{\tilde \varphi}_m)/\partial \varphi_p=-2E_J\cos\pi f_\alpha\sin\varphi_p
+{\tilde E}_J \cos{\tilde \varphi}_m\sin(\varphi_p/2)=0$  we have
\begin{eqnarray}
\label{phi}
{\tilde \varphi}_{m,i}=0 ~~{\rm and}~~ \cos\left(\frac{\varphi_{p,i}}{2}\right)=\frac{{\tilde E}_J}{4E_J\cos(\pi f_\alpha)}
\end{eqnarray}
with $i=1,2$. The trapping loop currents of qubit states,
$|\downarrow\rangle$ and $|\uparrow\rangle$, are shown in Figs.
\ref{scheme} (b) and (c), where they are in opposite directions
with each other while the directions of $\alpha$-junction loop are
the same. In the qubit state $|\uparrow\rangle
(|\downarrow\rangle)$,  we calculate the trapping loop current
 $I'_i=-(n_cAq_c/m_c)\hbar k'_i$ with $k'_i$ in Eq. (\ref{k'i}).
Since  $(n_cAq_c/m_c)\hbar(2\pi L'_k/l')=\Phi_0$, we can calculate numerically the reduced dimensionless
current as $I'_1 L_{\rm eff}/\Phi_0=I'_2 L_{\rm eff}/\Phi_0=\pm 0.00123$ for $f_1=f_2=0.94$, $L_M/(L_k+L_g)=0.4$,
$f_\alpha=0.2$ and ${\tilde E}_J /E_J= 2.0$ with   $m=0$, $n=1$ and $m'=-2$.
Here the integer $m'$ is determined by minimizing the potential energy ( see the details in appendix A).
The $\alpha$-junction loop current is given by $I_\alpha=(I_1-I_2)/2$ with $I_i=-(2\pi E_J/\Phi_0)\sin\varphi_i$
and thus $I_iL_{\rm eff}/\Phi_0=-(\pi/2)(4L_{\rm eff}E_J/\Phi^2_0)\sin\varphi_i$.
For the parameter value of $\Phi^2_0/4L_{\rm eff}E_J=1000$ with $L_{\rm eff}$=15pH and $E_J/h$=200GHz
we obtain $I_\alpha L_{\rm eff}/\Phi_0=0.00022$, i. e., $|I'_i|=$ 170nA and $I_\alpha= 30$nA.

The qubit energy gap $\Delta$ of the GFQ can be controlled by the ratio ${\tilde E}_J/E_J$ as well as the magnetic flux
$f_\alpha$ threading the $\alpha$-junction loop. We calculate numerically the qubit energy gap $\Delta=2t_q$
with $t_q$  being the tunnelling amplitude across the one-dimensional
potential well in Fig. \ref{contourwell}(b) \cite{Kim03}. We can
identify the parameter regime for a specific qubit energy gap
$\Delta$  with the ratio of the Josephson coupling energy to the
charging energy $E_J/E_C=40$ \cite{Wal}. In order to obtain
$\Delta/h\sim$1GHz, for example, we adjust $f_\alpha= 0.2$ with
 ${\tilde E}_J/E_J =2.0$.
Actually,  as $f_\alpha$ or ${\tilde E}_J/E_J$ increases, the barrier of double well potential decreases and thus $\Delta$ increases.



For   $I_0\neq 0$  the effective potential $V(\varphi_p,{\tilde \varphi}_m)= 2E_J\cos(\pi f_\alpha)\cos\varphi_p
-2{\tilde E}_J\cos(\varphi_p/2)\cos{\tilde \varphi}_m+(\Phi_0 I_0/2\pi)\varphi_p$
becomes tilted as shown in Fig. \ref{contourwell}(b). The energy levels can be represented as
$\left(E_{\downarrow}-\Phi_0I_0\alpha/2\pi\right)|\downarrow\rangle\langle\downarrow|
+\left (E_{\uparrow}+\Phi_0I_0 \alpha/2\pi\right)|\uparrow\rangle\langle\uparrow|,$
and thus the tight binding  Hamiltonian can be written as
\begin{eqnarray}
\label{H}
\fl H \!\!=\! E_{\downarrow} |\downarrow\rangle\langle\downarrow|\!+\!E_{\uparrow}|\uparrow\rangle\langle\uparrow|
\!-\! t_q(|\downarrow\rangle\langle\uparrow|\!+\!|\uparrow\rangle\langle\downarrow|)
 -\!\frac{\Phi_0I_0}{2\pi}\alpha (|\downarrow\rangle\langle\downarrow| \!-\! |\uparrow\rangle\langle\uparrow|),
\end{eqnarray}
where from Eq. (\ref{phi})  $\alpha=|\varphi_{p1}|=|\varphi_{p2}|$ is given by
\begin{eqnarray}
\label{alpha}
\alpha =2\cos^{-1}\left({\tilde E}_J/4E_J\cos(\pi f_\alpha)\right).
\end{eqnarray}

Here we consider that an ac bias current $I_0=-I_b\sin\omega t$ is
applied,
and  introduce a coordinate transformation such as
$|0\rangle=(|\downarrow\rangle+|\uparrow\rangle)/\sqrt{2}$ and
$|1\rangle=(|\downarrow\rangle-|\uparrow\rangle)/\sqrt{2}$.
For the degenerate case, $E_\downarrow=E_\uparrow$,
the Hamiltonian in Eq. (\ref{H}) can be transformed to
${\cal H}=(\Delta/2)\sigma_z+g\sin\omega t\sigma_x$
in the basis of $\{|0\rangle,|1\rangle\}$.
The coupling constant $g=(\Phi_0I_b/2\pi)\alpha$ also
depends on  ${\tilde E}_J/E_J$ and $f_\alpha$ such that
\begin{eqnarray}
\label{g}
g=\frac{\Phi_0 I_b}{\pi}\cos^{-1}\left(\frac{{\tilde E}_J}{4E_J\cos(\pi f_\alpha)}\right).
\end{eqnarray}
In Fig. \ref{contourwell}(c) we show the coupling strength $g$ as a function of $f_\alpha$ for several values
of ${\tilde E}_J/E_J$. In this study we set ${\tilde E}_J/E_J=2.0$ and $f_\alpha=0.2$ so that
$g/\Phi_0I_b\approx$0.3.

\begin{figure}[b]
\vspace{0cm}
\hspace{0cm}
\includegraphics[width=1.1\linewidth]{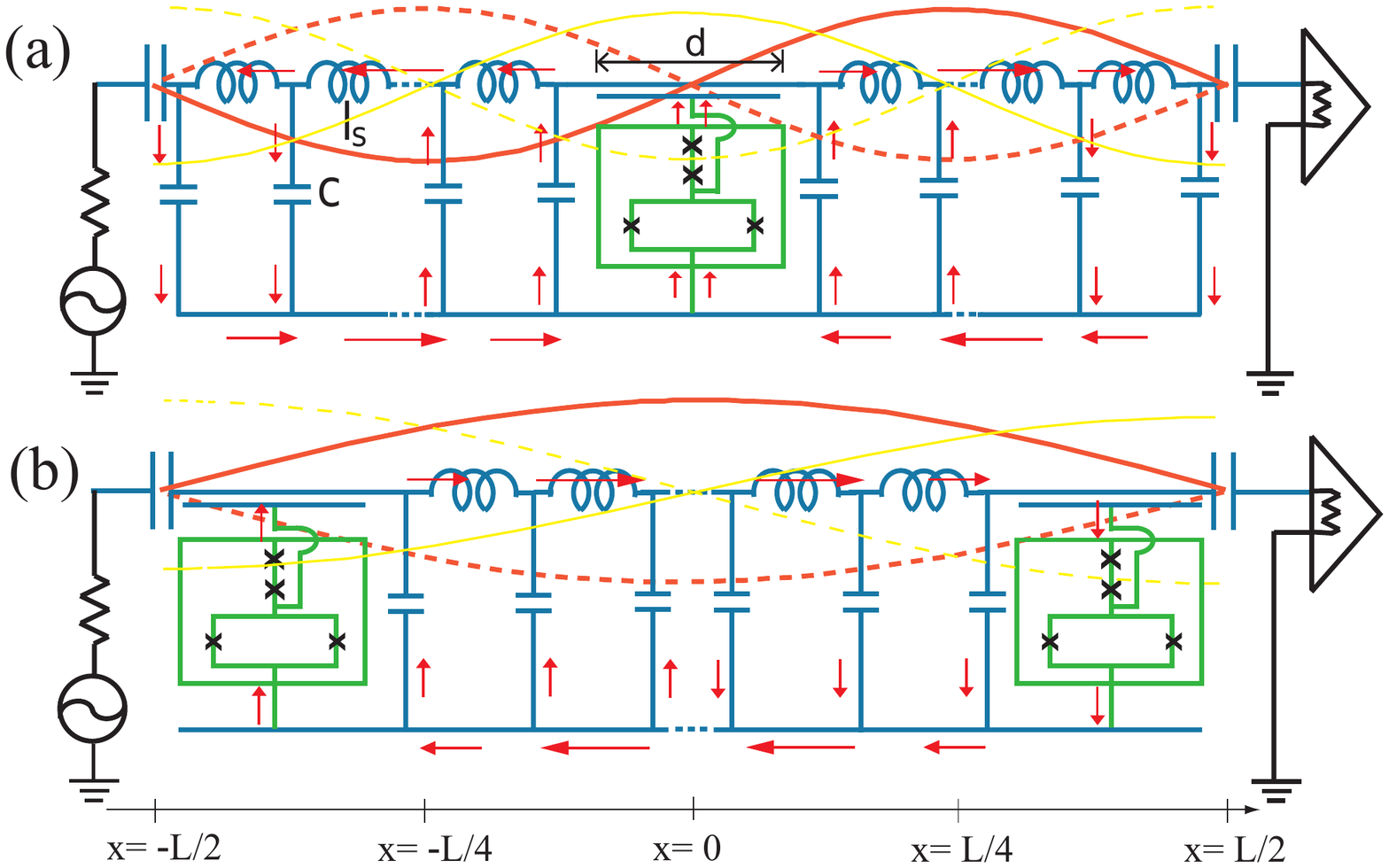}
\vspace{-5cm}
\caption{(a) A GFQ coupled with an current mode (orange)  and a voltage mode (yellow)
of the transmission line resonator of the circuit-QED architecture through a large capacitance
with width $d$, where arrows show the current directions and amplitudes.
(b) Two GFQs at  the ends of the resonator are coupled with each other through the current mode of resonator.
}
\label{scale}
\end{figure}

\section{Circuit QED with gradiometric flux qubits}

In Fig. \ref{scale}(a) we present a scheme for the GFQ in a circuit-QED architecture,
where a GFQ is coupled with the resonator through a large capacitance with width $d$.
The state of transmission line resonator of the circuit-QED architecture  can be described in terms of
the electric potential and current modes. The Lagrangian of the resonators is represented as
${\cal L}(\theta,{\dot\theta};t)=\int^{\frac{L}{2}}_{-\frac{L}{2}}
\left[(l_s/2)\partial^2_t\theta(x,t)-(1/2c)\partial^2_x\theta(x,t)\right]dx,$
where $l_s$ and $c$ are the inductance and the capacitance per unit
length of the uniform transmission line resonator, respectively, and
$\theta(x,t)=\int^x_{-L/2}dx'Q(x',t)$ with the linear charge density $Q(x,t)$ \cite{Blais04}.
As shown in Fig. \ref{scale}(a) the charge neutrality imposes the boundary condition,
$\theta(-L/2,t)=\theta(L/2,t)=0$, which allows the normal mode expansion such as
$\theta(x,t)=\sqrt{\frac{2}{L}}\sum_{n}\left(q_{2n-1}(t)\cos\frac{(2n-1)\pi x}{L}+
q_{2n}(t)\sin\frac{2n\pi x}{L}\right)$
with a positive integer  $n$. Here, the electric potential can be
represented as $V(x,t)=(1/c)\partial\theta(x,t)/\partial x$ and
the current $I(x,t)=\partial\theta (x,t)/\partial t$. The
Hamiltonian, then, can be derived from the Lagrangian
 as ${\cal H}= \sum_n[(l_s/2){\dot q}^2_n +(1/2c)(n\pi/L)^2q^2_n]$
which can be diagonalized as ${\cal H}=\sum_n\hbar\omega_n(a^\dagger_na_n+1/2)$
with the representation
${\hat q}_n(t)=\sqrt{\hbar\omega_n c/2}(L/n\pi)(a_n(t)+a_n^\dagger(t))$,
${\hat p}_n(t)=-i\sqrt{\hbar\omega_n l_s/2}(a_n(t)-a_n^\dagger(t))$ and $p_n=l_s{\dot q}_n$.
Here, $a^\dagger_n$ and $a_n$  are bosonic operators satisfying $[a_m,a^\dagger_n]=\delta_{m,n}$
and $\omega_n=n\pi v/L$ with $v=1/\sqrt{l_sc}$.

In  Fig. \ref{scale}(a) the current mode displayed as orange line can be obtained by
$I(x,t)=\partial\theta (x,t)/\partial t$  for $n=2$ as
\begin{eqnarray}
\label{I2}
{\hat I}_2(x,t)&=&-i\sqrt{\frac{\hbar\omega_2}{l_sL}}\sin\frac{2\pi x}{L}(a_2(t)-a^\dagger_2(t)).
\end{eqnarray}
From Eqs. (\ref{H}) and (\ref{I2}) we can have the Hamiltonian for the combined system of the resonator and the GFQ such as
\begin{eqnarray}
\label{H2}
\fl H \!=\! \hbar\omega_2\!\left(\!a^\dagger_2a_2\!+\!\frac12\!\right)\!\!
+\! E_{\downarrow}|\!\downarrow\rangle\langle\downarrow\!|\!+\!
E_{\uparrow}|\!\uparrow\rangle\langle\uparrow\!|\!-\! t_q(|\!\downarrow\rangle\langle\uparrow\!|\!
+\!|\!\uparrow\rangle\langle\downarrow\!|)
 \!+\!ig(|\!\downarrow\rangle\langle\downarrow\!| \!-\!|\!\uparrow\rangle\langle\uparrow\!|)
 (a_2\!-\!a^\dagger_2). \nonumber\\
\end{eqnarray}
If we assume a constant capacitance $c$ per unit length in Fig. \ref{scale}(a),
the bias current ${\hat I}_0=\int^{d/2}_{-d/2}{\hat{\dot q}}(x,t)dx$ with 
${\hat{\dot q}}(x,t)=\partial {\hat I}_2(x,t)/\partial x$ can be represented as 
${\hat I}_0={\hat I}_2(d/2,t)-{\hat I}_2(-d/2,t)=-2i\sqrt{\hbar \omega_2/l_s L} \sin(\pi d/L)
 (a_2 (t)-a^\dagger_2(t))$, 
and thus the amplitude $I_b$ of the bias current is given by  
$I_b=2\sqrt{\hbar \omega_2/l_s L} \sin(\pi d/L)$ 
which has a finite value for $d\neq 0$. 
In fact, however, the capacitance density should be larger around the large capacitor.  This general situation has been considered to calculate the bias current amplitude $I_b=\sqrt{\hbar\omega_2/l_sL}\delta$  
with a parameter $\delta$  depending on the parameters of capacitor
between the qubit and resonator \cite{Kim15}. 
The coupling strength  $g$, then, is given by  $g= \alpha(\Phi_0/2\pi)\delta\sqrt{(\hbar\omega_2/l_sL)}$
with  $\alpha$ in Eq. (\ref{alpha}), where
the coupling strength can be  sufficiently strong,  $g\sim \hbar\omega_1$,
due to the large capacitance between the qubit and resonator.
%
%
For the degenerate case that $E_\downarrow=E_\uparrow$
this Hamiltonian can be represented in the basis of $\{|0\rangle,
|1\rangle \}$ as
${\tilde H}= \hbar\omega_2a^\dagger_2a_2+(\Delta/2)\sigma_z+ig\sigma_x(a_2-a^\dagger_2)$,
and further transformed to
${\tilde H}_{\rm RWA}= \hbar\omega_2a^\dagger_2a_2+(\Delta/2)\sigma_z-ig(a^\dagger_2\sigma_- - \sigma_+a_2)$
in the rotating wave approximation.

Two GFQs can be coupled to perform  the two-qubit gate operations. As shown in Fig. \ref{scale}(b)
two GFQs are coupled through  the current mode, $I_1(x,t)$ for $n=1$, of the  resonator as
\begin{eqnarray}
{\hat I}_1(x,t)&=&-i\sqrt{\frac{\hbar\omega_1}{l_sL}}\cos\frac{\pi x}{L}(a_1(t)-a^\dagger_1(t)),
\end{eqnarray}
from which    the coupling strength is given by  $g_{l(r)}= \alpha_{l(r)}(\Phi_0/2\pi)\delta\sqrt{(\hbar\omega_1/lL)}$
for left(right) qubit  with $\alpha_{l(r)}$ being   $\alpha$ in Eq. (\ref{alpha}).
Then  the Hamiltonian for two coupled qubits in the dispersive regime,  $\Delta'_j=\Delta_j-\omega_1 \gg g_j$,
is written as \cite{Kim15,Blais07}
\begin{eqnarray}
{\cal H}_{2qubit}=\omega_1 a^\dagger_1 a_1\!+\!\sum_{j=l,r}\frac{{\tilde \Delta}_j}{2}  \sigma_{z_j}
\!+\!\frac12\left(\frac{1}{\Delta'_l}\!+\!\frac{1}{\Delta'_r}\right)g_l g_r
(\sigma_{-_l}\sigma_{+_r}\!+\!\sigma_{+_l}\sigma_{-_r}),
\end{eqnarray}
where ${\tilde \Delta}_j=\Delta_j+g^2_j/\Delta'_j$ with $\Delta_{l(r)}$ being the energy gap for left(right) qubit.



\section{Conclusions}
The GFQ is insensitive to the magnetic flux fluctuations due to the symmetry of design,
but at the same time it is difficult to manipulate the GFQ states by an external magnetic flux.
In this study, thus, we introduced a scheme for controlling the GFQ by an ac bias current.
By deriving exactly the effective Lagrangian of the system we showed that
the phase variables of the $\alpha$-junction loop  in Fig. \ref{scheme}(a)
of the GFQ  become coupled with the ac bias current $I_0$ as in
the last term of Eq. (\ref{V}). However, if we try to couple all phases,
$\varphi_i (i=1\sim 4)$,  with $I'_0$  in Fig. \ref{scheme}(a), the coupling term becomes constant,
which means that the GFQ do not respond to the external current  $I'_0$
and thus to the charge fluctuations with the length scale larger than the size of GFQ.
Therefore, the present GFQ scheme is robust against charge fluctuations as well as magnetic fluctuations,
which ensures a long coherence time for NISQ computing.

Further we obtained an analytic expression for the coupling strength $g$
between the bias current and the phase variables of the GFQ,
which is sufficiently strong for the quantum gate operations.
We also introduced a circuit-QED scheme involving the GFQs
to provide a scheme for performing  the single- and two-qubit operations.

\ack
This research was  supported by Basic Science Research Program through the National Research Foundation of Korea(NRF)
funded by the Ministry of Education(2019R1I1A1A01061274), 2021 Hongik University Research Fund,
and  Korea Institute for Advanced Study(KIAS) grant funded by the Korea government.


\appendix


\section{A constant term in effective potential  $U_{\rm eff}$ }


\begin{figure}[b]
\vspace{-2cm}
\centering
\includegraphics[width=1.0\linewidth]{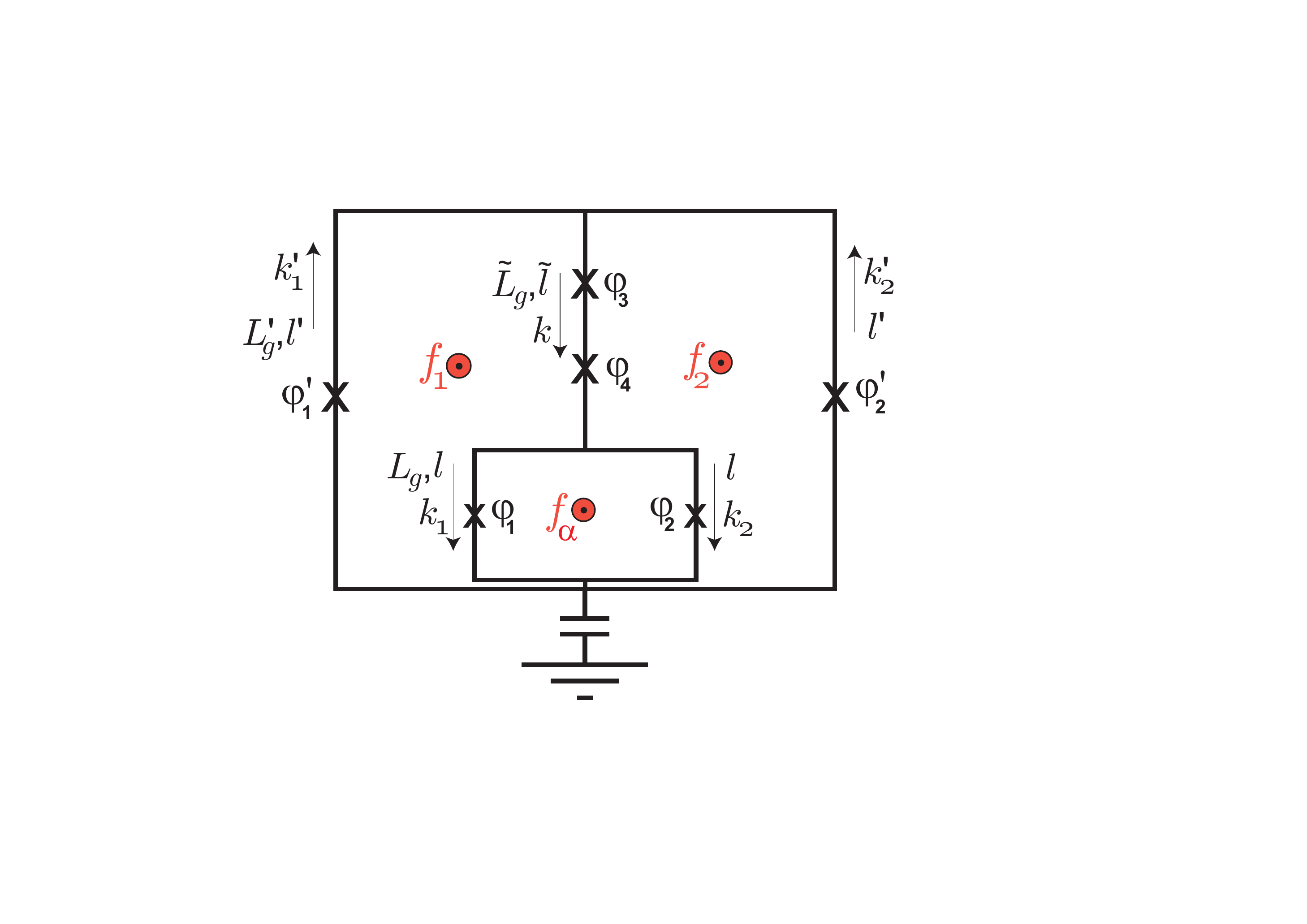}
\vspace{-3cm} \caption{  Two Josephson junctions with phase
differences $\varphi'_i$ are additionally introduced in
the left and right trapping loops of a gradiometric flux qubit.}
\label{schemeSI}
\end{figure}

We  consider a GFQ scheme different from that in Fig.
\ref{scheme}(a). In Fig. \ref{schemeSI} two Josephson junctions
with phase differences, $\varphi'_1$ and $\varphi'_2$,  in
trapping loops are additionally introduced. The boundary
conditions then become
\begin{eqnarray}
\label{bc1-a}
&-&k_1 l-k'_1 l'-k{\tilde l}-\varphi_1-\varphi_3-\varphi_4+\varphi'_1=2\pi(n_1+f_1+f_{1,{\rm ind}})\\
\label{bc2-a}
&&k_2 l+k'_2 l'+k{\tilde l}+\varphi_2+\varphi_3+\varphi_4-\varphi'_2=2\pi(n_2+f_2+f_{2,{\rm ind}})\\
\label{bc3-a}
&&k_1l-k_2l+\varphi_1-\varphi_2=2\pi(n+f_\alpha+f_{\alpha,{\rm ind}}),
\end{eqnarray}
and with the induced fluxes, $f_{\rm ind,i}=\Phi_{\rm
ind,i}/\Phi_0$, in the main manuscript these conditions  can be
represented as
\begin{eqnarray}
\label{lbc-a}
\fl &\!-\!&\!\!\left(\!1\!\!+\!\!\frac{L_g}{L_K}\!\right)\!\!k_1 l\!-\!\!\left(\!1\!\!+\!\!\frac{L'_g}{L'_K}\right)\!\!k'_1 l'\!
\!-\!\!\left(\!1\!\!+\!\!\frac{{\tilde L}_g}{{\tilde L}_K}\right)\!\!k {\tilde l}\!-\!\!\frac{L_M'}{L'_K}k'_2 l'\!\!-\!\!\frac{L_M}{L_K}k_2 l
\!=\!2\pi\!\left(\!\!n_1\!\!+\!\!f_1\!\!+\!\!\frac{\varphi_1\!+\!\varphi_3\!+\!\varphi_4\!-\!\varphi'_1}{2\pi}\right)\nonumber\\
\fl \\
\label{rbc-a}
\fl &&\!\!\left(\!1\!\!+\!\!\frac{L_g}{L_K}\right)\!\!k_2 l\!+\!\!\left(\!1\!\!+\!\!\frac{L'_g}{L'_K}\right)\!\!k'_2 l'\!
\!+\!\!\left(\!1\!\!+\!\!\frac{{\tilde L}_g}{{\tilde L}_K}\right)\!\!k {\tilde l}\!+\!\!\frac{L_M'}{L'_K}k'_1 l'\!\!+\!\!\frac{L_M}{L_K}k_1 l
\!=\!2\pi\!\!\left(\!n_2\!+\!f_2\!-\!\frac{\varphi_2\!+\!\varphi_3\!+\!\varphi_4\!-\!\varphi'_2}{2\pi}\right)\nonumber\\
\fl \\
\label{cbc-a}
\fl &&\!\!\left(\!1\!+\!\frac{L_g}{L_K}\right)\!(k_1\!-\!k_2)l
\!=\!2\pi\!\left(\!n\!+\!f_\alpha\!-\!\frac{\varphi_1\!-\!\varphi_2}{2\pi}\right).
\end{eqnarray}

With the current conservation conditions at nodes, $k=k'_1+k'_2$ and $k=k_1+k_2$, we obtain
\begin{eqnarray}
\fl k_{1,2}\!&=&\! \frac{2\pi L_K}{l}\!\!\left[\frac{1}{2L_{\rm eff}}\!\!\left(\!\!m\!+\!f_2\!\!-\!\!f_1\!\!
-\!\!\frac{\varphi_1\!+\!\varphi_2\!+\!2\varphi_3\!+\!2\varphi_4\!\!-\!\!\varphi'_1\!\!-\!\!\varphi'_2}{2\pi}\!\right)\!\!\pm\!\!\frac{1}{2(L_K\!\!+\!\!L_g)}\left(n\!\!+\!\!f_\alpha\!\!-\!\!\frac{\varphi_1\!\!-\!\!\varphi_2}{2\pi}\right)\right]\nonumber\\
\fl\\
\fl k'_{1,2}\!&=&\!\frac{2\pi L'_K}{l'}\!\!\left[\frac{1}{2L_{\rm eff}}\!\!\left(\!\!m\!+\!f_2\!\!-\!\!f_1\!\!
-\!\!\frac{\varphi_1\!+\!\varphi_2\!+\!2\varphi_3\!+\!2\varphi_4-\!\!\varphi'_1\!\!-\!\!\varphi'_2}{2\pi}\!\right)\right.\nonumber\\
\fl &\mp&\left.\frac{1}{2(L'_K\!\!+\!\!L'_g-L_M')}\left(m'\!+\!f_1\!+\!f_2\!
+\left(\!1\!-\!\frac{L_M}{L_K\!+\!L_g}\right)\!f_\alpha\!-\!\frac{\varphi'_1\!\!-\!\!\varphi'_2}{2\pi}\right)\right] \\
\fl k\!&=&\!\frac{2\pi {\tilde L}_K}{{\tilde l}L_{\rm eff}}
\!\!\left(m+f_2-f_1-\frac{\varphi_1+\varphi_2+2\varphi_3+2\varphi_4-\varphi'_1-\varphi'_2}{2\pi}\right),
\end{eqnarray}
where $L_{\rm eff}= L_K+L_g+L'_K+L'_g+2({\tilde L}_K+{\tilde L}_g)+L_M+L_M'$,  $m=n_2-n_1$ and $m'=n_1+n_2$.

In order to satisfy the quantum Kirchhoff relation,
$(\Phi^2_0/2\pi L_K)(l/2\pi)k_i-E_{J}\sin\phi_i=-\partial U_{\rm eff}/\partial\phi_i$,
the effective potential $U_{\rm eff}$ should be
\begin{eqnarray}
\label{Ueff-a}
\fl U_{\rm eff}(\{\varphi_i,\varphi'_i\})&\!=\!\!\!\!&\!\!\frac{\Phi^2_0}{4L_{\rm eff}}
\!\!\left(\!\!m\!+\!\!f_2\!-\!\!f_1\!-
\!\frac{\varphi_1\!+\!\varphi_2\!+\!2\varphi_3\!+\!2\varphi_4\!-\!\varphi'_1\!-\!\varphi'_2}{2\pi}\!\right)^2
\!\!\!\!+\!\frac{\Phi^2_0}{4(L_K\!+\!L_g)}
\!\!\left(\!\!n\!+\!f_\alpha\!-\!\!\frac{\varphi_1\!-\!\varphi_2}{2\pi}\!\right)^2\nonumber\\
\fl &+&\!\frac{\Phi^2_0}{4(L'_K\!+\!L'_g)}\!\left[\!m'\!+\!f_1\!+\!f_2\!
+\!\left(\!1\!-\!\frac{L_M}{L_K\!+\!L_g}\right)\!f_\alpha\!-\!\frac{\varphi'_1\!-\!\varphi'_2}{2\pi}\right]^2
\!\!-\!\sum^4_{i=1}E_{Ji}\cos\varphi_i.
\end{eqnarray}
Here,  we consider the limit that $\varphi'_1$ and $\varphi'_2$
diminish to zero, which means that virtually there is no  effect
of the additionally introduced two Josephson junctions. Then the
boundary conditions in Eqs. (\ref{bc1-a})-(\ref{bc3-a}) reduce to
those in Eqs. (\ref{bc1})-(\ref{bc3}), but the effective potential
in Eq. (\ref{Ueff-a}) has an additional  constant term,
\begin{eqnarray}
\frac{\Phi^2_0}{4(L'_K+L'_g)}\left[m'+f_1+f_2+\!\left(\!1\!-\!\frac{L_M}{L_K\!+\!L_g}\right)f_\alpha\right]^2,
\end{eqnarray}
which should have been included in the effective potential in Eq.
(\ref{Ueff}). Actually this constant term cannot be obtained in
the effective potential for the GFQ in Fig. \ref{scheme}(a) from
the quantum Kirchhoff relation in Eq. (\ref{QK}). But in order to
calculate the current $I'$ with Eq. (\ref{k'i}) we need to
minimize this term to determine the value of integer $m'$.


\end{document}